# Relational Approach for Shortest Path Discovery over Large Graphs

Jun Gao† Ruoming Jin§ Jiashuai Zhou† Jeffrey Xu Yu‡ Xiao Jiang† Tengjiao Wang†
† Key Laboratory of High Confidence Software Technologies, EECS, Peking University
§ Department of Computer Science, Kent State University
‡ Department of Systems Engineering & Engineering Management, Chinese University of Hong Kong
{gaojun,zjiash,jiangxiao,tjwang}@pku.edu.cn, jin@cs.kent.edu, yu@se.cuhk.edu.hk

## ABSTRACT

With the rapid growth of large graphs, we cannot assume that graphs can still be fully loaded into memory, thus the disk-based graph operation is inevitable. In this paper, we take the shortest path discovery as an example to investigate the technique issues when leveraging existing infrastructure of relational database (RDB) in the graph data management.

Based on the observation that a variety of graph search queries can be implemented by iterative operations including selecting frontier nodes from visited nodes, making expansion from the selected frontier nodes, and merging the expanded nodes into the visited ones, we introduce a relational *FEM* framework with three corresponding operators to implement graph search tasks in the RDB context. We show new features such as *window function* and *merge statement* introduced by recent SQL standards can not only simplify the expression but also improve the performance of the *FEM* framework. In addition, we propose two optimization strategies specific to shortest path discovery inside the *FEM* framework. First, we take a bi-directional set Dijkstra's algorithm in the path finding. The bi-directional strategy can reduce the search space, and set Dijkstra's algorithm finds the shortest path in a set-at-a-time fashion. Second, we introduce an index named SegTable to preserve the local shortest segments, and exploit SegTable to further improve the performance. The final extensive experimental results illustrate our relational approach with the optimization strategies achieves high scalability and performance.

## 1. INTRODUCTION

The rapid growth of graph data raises significant challenges to graph management. Nowadays, the graph data are used in numerous applications, *e.g.*, web graphs, social networks, ontology graphs, transportation networks. These graphs are always exceedingly large and keep growing at a fast rate. When the graph cannot fit into the memory, the existing in-memory graph operations have to be re-examined. We need the buffer mechanism to load graph blocks on demand, in which case the I/O cost becomes the key factor in the evaluation cost. We also require flexible approaches to access nodes and edges. Moreover, the stability of graph management system is another concern.

Graph search is an important and primitive requirement on the graph management, which seeks a sub-graph(s) meeting the specific purposes, such as the shortest path between two nodes [12], the minimal spanning tree [20], the path for traveling salesman [9], and the like. This paper focuses on the shortest path discovery problem due to two reasons. First, the shortest path discovery plays a key role in many applications. For example, the shortest path discovery in a social network between two individuals reveals how their relationship is built [21]. Second, the shortest path discovery is a representative graph search query, which has a similar evaluation pattern to other queries [20, 9].

The current disk-based methods face limitations to support the general graph search queries. The index on the external memory for shortest path discovery has been designed, but the method is restricted to planar graphs [8]. We notice that the MapReduce framework [16, 3] and its open source implementation Hadoop [1] can process large graphs stored in the distributed file system over a cluster of computers. However, due to the lack of schema and index mechanism, it is difficult to access graphs in a flexible way. In addition, it is expensive to support the dynamic changes of the original graph, at least in the current Hadoop distributed file system [1]. Some other graph operations, such as minimum-cut [5] and clique computation [15], can be evaluated on the partially loaded graph when the quality of results is assured theoretically or the approximate results are allowed [15, 5]. However, it is difficult to extend these methods to other graph operations.

RDB provides a promising infrastructure to manage large graphs. After more than 40 years' development, RDB is mature enough and plays a key role in the information system. We can notice that RDB and the graph data management have many overlapping functionalities, including the data storage, the data buffer, the access methods, and the like. Indeed, RDB provides a basic support to the graph storage and the flexible access to the graph. In addition, RDB has already shown its flexibility in managing other complex data types and supporting novel applications. For example, RDB can support the XML data management [17, 13], and can be used in reachability query and breadth-first-search (BFS) in the graph management [23, 22]. The statistical data analysis and data mining over RDB have also been studied in [6, 7].

However, it is a challenging task to translate the graph search methods and improve the performance in the RDB context due to the semantic mismatch between relational operation and graph operation. The data processing in the graph search is always complex. We may need do logic and arithmetic computation, make choices based on aggregate values, and record necessary information in the searching [12, 20, 9]. At the same time, the graph and running time data have to be stored in the tables in RDB first, and only restricted operations, including, projection, selection, join, aggregation, etc, are allowed to perform on tables [14].

In this paper, we investigate the techniques used in the shortest path discovery in the RDB context. We wish our method can not only deliver benefits to the shortest path computation, but also shed lights on the SQL implementation for other graph search queries and graph analysis. Specifically, the contributions of this paper can be summarized as follows:

- Based on the observation that many graph search queries can be evaluated by iterative operations including selecting frontier nodes from visited nodes, making expansion from the frontier nodes, and merging the expanded nodes into the visited nodes, we introduce a generic graph processing framework *FEM* with three key operators $F$, $E$ and $M$-operator. We find new features, such as *window function* and *merge statement* introduced by recent SQL standards can be used to simplify the expression and improve the performance. We also show how the shortest path is discovered using the *FEM* framework. (Section 3)

- We propose two optimizations specific to the shortest path discovery inside the *FEM* framework. First, we take a bi-directional set Dijkstra's method. The bi-directional searching can find the path with less search space. Set Dijkstra's method allows all nodes with the same minimal distance to be expanded in one operation, which is more suitable in RDB context. Second, we introduce an index named SegTable to preserve local shortest segments, and exploit SegTable to further improve the performance by balancing reduction of search space and promotion of set-at-a-time query evaluation. (Section 4)

- We conduct extensive experiments over synthetic and real-life data. The results show that our relational approach has high scalability in handling large graphs. We also see that the new features of SQL standard, as well as our optimizations can improve the performance significantly. (Section 5)

## 2. PRELIMINARY

In this section, we first give the related notations and relational representation of graph, and then show new SQL features used in our path finding method.

### 2.1 Graph Notations

This paper studies weighted (directed or undirected) graphs. Let $G = (V, E)$ be a graph, where $V$ is a node set and $E$ is an edge set. Each node $v \in V$ has a unique node identifer. Each edge $e \in E$ is represented by $e = (u, v)$, $u, v \in V$. For an edge $e = (u, v)$, $e$ is $u's$ outgoing edge and $v's$ incoming edge. Each edge $(u, v) \in E$ has its non-negative weight $w(u, v)$. The minimal edge weight in a graph is denoted by $w_{min}$. A path $p = u_0 \leadsto u_x$ from $u_0$ to $u_x$ is a sequence of edges $(u_0, u_1), (u_1, u_2), \ldots, (u_{x-1}, u_x)$, where $e_i = (u_i, u_{i+1}) \in E$ ($0 \leq i < x$). The length of a path $p$, $len(p)$, is the sum of the weights of its constituent edges. The shortest distance from $u$ and $v$ is denoted by $\delta(u, v)$.

We can store a graph into relational tables easily, which are illustrated in Figure 1. Let $G = (V, E)$ be a graph. We use *TNodes* table to represent nodes $V$. The *nid* in the table is for the node's unique identifer. We use *TEdges* table to store edges $E$. For an edge $(u, v)$, the identifiers of nodes $u$ and $v$, as well as the weight of the edge, are recorded by *fid*, *tid* and *cost* field respectively.

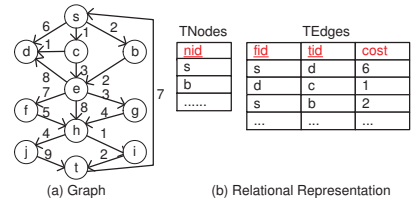

**Figure 1: Relational Representation of Graph**

### 2.2 SQL Features

In this paper, in addition to utilize standard features in SQL statements, we leverage two new SQL features, window function and merge statement, which are now supported by Oracle, DB2, and SQL server. These features are the "short-cut" for combinations of some basic relational operators. In addition, they can improve the performance of the statements.

**Listing 1: Syntax for New Features**
```
1:Window function
<WindowFunction> :: = <Aggregation>
  Over ( [ Partition By <expr>, ... ]
    [ Order By <expression> ] )
2: Merge statement
Merge Into tablename Using table On (condition)
When Matched Then
 Update Set col1 = val1 [, col2 = val2 ...]
When Not Matched Then
 Insert (col1[,col2 ...]) Values (val1[, val2 ...])
```

The window function [1], introduced by SQL 2003, returns aggregate results. Compared to traditional aggregate functions which return one value for one tuple set, window function obtains the aggregate result for each tuple in the set, and then the non-aggregate attributes are allowed to be along with the aggregate ones in the *select* clause, even if they are not in the *group by* clause. In addition, window function supports the aggregate functions related to the tuple position in a sorted tuple sub-set, such as *rank*, *row_num*, and the like. The syntax for window function can be described in Listing 1(1).

The merge statement [2] adds new tuples or updates the existing ones in a target table from a source ( SQL results, a view or a table). It is officially introduced by SQL 2008, but is supported earlier by different database vendors due to the need in loading data into data warehouse. The main part of merge statement specifies actions, like *insert*,*delete* and *update* under different relationships between the source and target table. Compared to the general-purpose one *update* statement followed by an *insert* statement, the merge statement is more specific, which indicates that it can be evaluated faster than two statements. The syntax for merge statement can be referred in Listing 1(2).

## 3. RELATIONAL FEM FRAMEWORK FOR SHORTEST PATH DISCOVERY

In this section, we introduce a generic graph search framework *FEM* and leverage it to realize the classical Dijkstra's shortest path discovery algorithm on RDB. In the next section, we will explore several new techniques to speed up the shortest path discovery under the relational *FEM* framework.

---
[1]http://en.wikipedia.org/wiki/SQL:2003
[2]http://en.wikipedia.org/wiki/Merge_(SQL)

## 3.1 A Generic Processing Framework for Graph Search

Graph search queries seek the sub-graph(s) meeting the specific requirements. For example, reachability query answers whether there exists a path between two given nodes [11]. The shortest path query locates the shortest path between two given nodes [19, 2]. The minimal spanning query returns a tree covering all nodes with the minimal sum of edge weights [20]. The salesman path query gets a shortest possible tour that visits each city exactly once [9]. The graph pattern match retrieves all sub-graphs, each of which is isomorphic to the given graph pattern [25].

Many graph search algorithms show a common pattern. Due to the large search space, they always utilize greedy ideas. In addition, we observe that most of these greedy algorithms can fit into a generic iterative processing structure. A visited node set $A^1$ is initialized first according to different purposes. Then an iterative searching starts. We illustrate the $k^{th}$ iteration in Figure 2. Let the **visited nodes** $A^k$ record all the nodes encountered in the graph search so far; the **frontier nodes** $F^k$ are selected from the visited nodes with the certain criteria (application dependent, $F^k \subseteq A^k$); the **expanded nodes** $E^k$ are the next visited nodes from the frontier nodes (generally neighboring subset of $F^k$); and the next visited ones $A^{k+1}$ are obtained by merging the newly expanded nodes $E^k$ and $A^k$. The iterations continue until the target sub-graph(s) can be discovered. We refer to such a generic processing structure as a **FEM** framework.

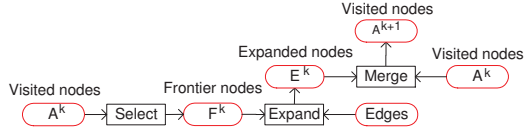

**Figure 2: Conversion Between Nodes in the $k^{th}$ Iteration**

Before we proceed to the details of implementing the relational operators for *FEM* framework, we first exploit it to briefly describe three representative graph search algorithms: *Dijkstra's shortest path algorithm*, *Prim's minimal spanning tree algorithm*, and *graph pattern matching*. In the Dijkstra's algorithm for shortest path discovery [12] using the *FEM* framework, each node is annotated with $d2s$ to record the distance from the source node $s$, and a flag $f$ indicating whether the node is finalized or not. We initially add the source node $s$ into visited nodes with $s.d2s = 0$ and $s.f = false$. We then start an iterative path expansion for the target node. In each iteration, we *select* a non-finalized node $u$ ($u.f = false$) from all visited nodes with the minimal $d2s$ as a frontier node, finalize $u$, *expand* $u$ (visiting its neighbors), and *merge* the newly expanded nodes into the visited nodes.

Another case is to construct a minimal spanning tree $T$ by Prim's algorithm [20]. Each node $u$ is represented by $u(p2s, w, f)$. Here $p2s$ is the parent of $u$, $w$ is the edge weigh from $u$ to $p2s$, if $u$ is in $T$. $f$ is a flag for whether $u$ is in $T$. The visited nodes are initialized by any node $u$ with $u.f = false$ and $u.w = 0$. In each iteration in building $T$, we *select* a node $u$ with $u.f = false$ and the minimal edge weight, add $u$ into $T$ by changing $u.f = true$, make further *expansion* from $u$, and *merge* the expanded nodes into the visited nodes. In the merge operation, the expanded nodes can be discarded directly if they have been included ($f = true$) in $T$. The iterations repeat until all nodes have been in $T$.

Graph pattern matching [25] is a more complex case. Let $q = (V_q, E_q)$ be a graph pattern over a graph $G = (V, E)$. Each query node in $V_q$ and data node in $V$ have labels. Initially, we start from any query node $q_u^0$, and obtain the visited nodes $\{(d_u^0)|d_u^0 \in V$ and $q_u^0 \in V_q$ have the same label $\}$. We then handle the other query nodes in $V_q$ iteratively. During processing the query node $q_u^k$, let the visited nodes $\{(d_u^0,\ldots,d_u^{k-1})|d_u^0 \in V,\ldots,d_u^{k-1} \in V$ have the same labels to $q_u^0 \in V_q,\ldots,q_u^{k-1} \in V_q$ correspondingly$\}$. We expand the visited nodes into $\{(d_u^0,\ldots,d_u^k)|d_u^0 \in V,\ldots,d_u^k \in V$ have the same labels to $q_u^0 \in V_q,\ldots,q_u^k \in V_q$ correspondingly, and the relationships among $d_u^0,\ldots,d_u^{k-1}$ and $d_u^k$ meet the requirements among $q_u^0,\ldots,q_u^{k-1}$ and $q_u^k\}$. After all query nodes have been handled, each element in the visited node set represents a sub-graph meeting the requirement.

Note that there are other operations besides the three basic operations (select, expand, and merge) in the graph search. For example, the recovery of the shortest path or the minimal spanning tree, and the termination detection, are also needed in the *FEM* framework. However, these operations are generally auxiliary under the *FEM* framework and their computational costs are quite minimal compared to the three main operations. The exploration on the shortest path search in the reminder of the paper will illustrate the details of these additional operations, and also demonstrate the key functionality of the three operations.

## 3.2 Relational FEM Operators for Shortest Path Discovery

To realize a graph search query using *FEM* framework on relational database, we need three operators which can be expressed by relational algebra: i) $F$-operator to *select* frontier nodes from the visited nodes; ii) $E$-operator to *expand* the frontier nodes; and iii) $M$-operator to *merge* the newly expanded nodes into the visited ones. As discussed above, the attributes on nodes and the operations may be different for different graph search tasks under the *FEM* framework. In the reminder of the paper, we will focus on efficiently realizing the shortest path discovery under *FEM* framework. The new techniques we developed for shortest paths in the *FEM* framework can be in general applied or extended to deal with other graph search algorithms.

Now, we start with Dijkstra's algorithm for the shortest path discovery as an example to describe its $F$, $E$ and $M$-operator. The visited nodes for the shortest path discovery can be represented by $A^k$, where $k$ is the number of iterations. Let $s$ be the source node. For each node $u \in A^k$, $u$ is represented by $(nid, d2s, p2s, f)$, where $nid$ is for the node identifier, $d2s$ is for the distance from $s$ to $u$, $p2s$ is for the predecessor node of $u$ in the path from $s$ to $u$, and $f$ for the sign indicating whether $u$ is finalized or not. Take the shortest path discovery from $s$ to $t$ in the graph in Figure 1 as an example. We add $s$ to $A^1$ first. $d, c, b$ will be added into $A^2$ next. Figure 3 shows the visited nodes, frontier nodes, and expanded nodes in the 2-nd iteration of the shortest path discovery.

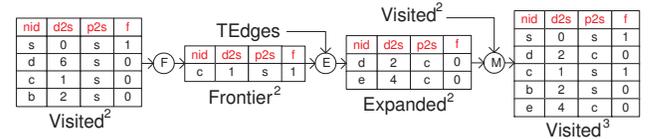

**Figure 3: $F$, $E$ and $M$-operator in the 2-nd Iteration**

Below we formally describe these operators based on the relational algebra.

DEFINITION 1. *F-operator returns frontier nodes $F^k$ from visited nodes $A^k$ in the $k^{th}$ expansion. $F^k \leftarrow \sigma_{nid=mid}A^k$.*

In Dijkstra's algorithm, we select the node with the minimal $d2s$ among all non-finalized nodes, and assign its identifier to $mid$. Take Figure 3 as an example, $mid$ is for node $c$ with the minimal $d2s = 1$. The computation of $mid$ can be done by an auxiliary operation before $F$-operator. In order to enhance the utility of our framework, we assume that there may be multiple frontier nodes after the $F$-operator. For example, the optimization strategy used in the next section may produce multiple frontier nodes with the revised predicate in the $F$-operator. After the $F$-operator, we use another auxiliary operation to adjust $f = 1$ for the node identified by $mid$.

DEFINITION 2. *E-operator* returns the expanded nodes $E^k$ based on frontier nodes $F^k$ and TEdges table in the $k^{th}$ expansion.

(1) $minCost(x, c) \leftarrow_x \mathcal{G}_{min(d2s+w)} \Pi_{(x,d2s,w)}(F^k(r, d2s, p2s, f) \bowtie TEdges(r, x, w));$

(2) $E_k \leftarrow \Pi_{(x,d2s+w,r,0)} \sigma_{c=d2s+w} minCost(x, c) \bowtie F^k(r, d2s, p2s, f) \bowtie TEdges(r, x, w);$

The $(r, x, w)$ in TEdges table is for the edge from $r$ to $x$ with the weight $w$.

In the $E$-operator, $minCost(x, c)$ preserves the newly visited nodes with the minimal distances. Since the expanded nodes may be reached by different paths, and only the ones with the minimal distance ($d2s + w$) are needed, we use an aggregate function to find them. Note that $minCost(x, c)$ contains the newly expanded nodes along with their costs, but lacks their parents $p2s$, which are required in the recovery of the full path. However, we cannot simply put the non-aggregate attribute $p2s$ into the *select* clause in $minCost(x, c)$, due to the constraint on the aggregation function in relational algebra. Thus, another join operation is required to find the parent node $p2s$ from *TEdges* under which the minimal distance can be achieved. Take Figure 3 as an example, we make the expansion from the frontier nodes $\{c\}$, and get the newly expanded nodes $d$ and $e$.

DEFINITION 3. *M-operator* returns visited nodes $A^{k+1}$ based on the expanded nodes $E^k$ and existing visited nodes $A^k$.

(1) $A^k \leftarrow A^k - \Pi_{x,d2s_1,p2s_1,f_1}(\sigma_{d2s_0 < d2s_1}$
$(E^k(x, d2s_0, p2s_0, f_0) \bowtie A^k(x, d2s_1, p2s_1, f_1)));$

(2) $E^k \leftarrow E^k - \Pi_{x,d2s_0,p2s_0,f_0}(\sigma_{d2s_0 > d2s_1}$
$(E^k(x, d2s_0, p2s_0, f_0) \bowtie A^k(x, d2s_1, p2s_1, f_1)));$

(3) $A^{k+1} \leftarrow A^k \cup E^k;$

In the $M$-operator, we remove the visited nodes from $A^k$, whose distances are larger than those of the corresponding newly expanded nodes in $E^k$. Then we remove the newly expanded nodes from $E^k$, whose distances are larger than those of the corresponding nodes in $A^k$. Finally, we union $A^k$ and $E^k$ to get the next visited node set $A^{k+1}$. Still use Figure 3 as an example, node $d$ can be replaced by the newly expanded one, and node $e$ is newly added into the visited nodes.

### 3.3 SQL Implementation for the Operators

Now, we discuss the SQL implementation for these operators. We use a table *TVisited* to store all visited nodes. The attributes $nid$, $d2s$, $p2s$, $f$ in *TVisited* have the same meanings above. Since $F$, $E$ and $M$-operator are based on relational algebra, we can use SQL to express them. However, the direct translation will result in a poor performance, especially for $E$ and $M$-operator. For example, $E$-operator is implemented by an aggregate function over the join results with *group by* clause. In addition, the location of the parent node $p2s$ of each expanded node still needs another join operation. Moreover, when there are multiple paths with the same minimal $d2s$ to the expanded node $x$, we have to keep only one due to the primary key constraint on $nid$ in *TVisited* table. As for $M$-operator, we have two different actions according to whether the expanded nodes have been in the visited nodes or not. In the SQL implementation, we might use an update statement followed by an insert statement with a *not exists* sub-query. Such a kind of expression is not only verbose but also inefficient.

**Listing 2: SQL in Path Finding**
```
1://Initialize TVisited with source node
Insert into TVisited(nid, d2s, p2s, f)
  values(s, 0, s, 0);
2://Locate the next node to be expanded
Select top 1 nid from TVisited where f=0
 and d2s=(select min(d2s) from TVisited where f=0);
3://E-operator in the k-th forward expansion
create view ek as
 select nid, p2s, cost
   from ( select out.tid, out.fid,
           out.cost+q.d2s, row_number() over
             (partition by out.tid order by
               out.cost+q.d2s asc) as rownum
           from TVisited q, TEdges out
           where q.nid=out.fid and q.nid= mid)
       tmp (nid, p2s, cost,rownum)
   where rownum=1
4://M-operator in the k-th forward expansion
Merge TVisited as target using ek as source
 on source.nid=target.nid
when matched and target.d2s>source.cost then
 update set d2s=source.cost, p2s=source.p2s, f=0
when not matched by target then
 insert (nid, d2s, p2s, f)
 values(source.nid, source.cost, source.p2s, 0);
```

Fortunately, we find the new features including window function and merge statement introduced by recent SQL standards can simplify the expression as well as improve the performance. Window function can return non-aggregate attributes along with aggregate results for the same tuple. As for our case, we can partition all occurrences of expanded nodes by the same identifer $nid$, sort them with their $d2s$ and select the tuple with the minimal $d2s$ by using an aggregate function $row\_number$. We see that the window function can avoid another extra join operation to locate the parent node $p2s$ of the currently expanded one. It also can handle the case when the node can be reached by multiple paths with the same minimal distances. As for the $M$-operator, we use one merge statement to combine two separated insert and update statements.

The SQL statement used in the path expansion is illustrated in Listing 2. The 1-st statement adds the source node $s$ into *TVisited* nodes initially, where $f$ is set to 0 (non-finalized). The $E$-operator can be expressed by the 3-rd statement. It makes a join operation between the frontier nodes specified by $nid = mid$ ($F$-operator) and *TEdges* table, where $mid$ is the identifer of the to-be-finalized node which is discovered by the 2-nd statement. The tuples with the minimal $d2s + c$ among multiple occurrences for the same node can be located by a window function $row\_number = 1$ over the sorted tuples. In the 4-th SQL statement for $M$-operator, we use one statement to merge the expanded nodes into *TVisited* table. When the newly expanded nodes are new to existing visited nodes in *TVisited*, we directly add them into *TVisited*. When the newly expanded nodes have smaller distances from the source node, the nodes in the *TVisited* are replaced by the newly expanded ones.

## 3.4 Shortest Path Discovery using FEM Framework

Here, we present Dijkstra's algorithm for shortest path discovery in Algorithm 1 as a case to show the functionality of the *FEM* framework. Besides the key SQLs in the path expansion, we also need auxiliary SQLs in Listing 3. Such an algorithm can run on the client side, which connects to the underlying RDB via JDBC or ODBC. In the running time, only few variables are kept on the client side, and the RDB carries out time-consuming tasks.

---

**Algorithm 1**: Shortest Path Discovery in *FEM* Framework

**Input**: source node $s$ and target node $t$, Graph $G = (V, E)$.
**Output**: The shortest path between $s$ and $t$.

1  Initialize *Tvisited* with the SQL in Listing 2(1);
2  **while** *true* **do**
3     Locate $mid$ for node $u's$ id with the SQL in Listing 2(2);
4     Expand path with the SQLs in Listing 2(3,4) with $mid$;
5     **if** *the number of affected tuples is 0* **then**
6        Break;
7     $u.f \leftarrow 1$ for the finalized node with the SQL in Listing 3(2) with $mid$;
8     **if** *there exists result for the SQL in Listing 3(1)* **then**
9        Break;
10  Iteratively find edges in the shortest path $p$ along $p2s$ link with the SQL in Listing 3(3);
11  Return $p$;

---

We initialize *TVisited* table first with the source node. We then start an iteration to find the shortest path from line 2 to line 9. $mid$ in line 3 is the node id for the to-be-finalized node, which can be located by an SQL on *TVisited* table. We then use $mid$ to compose the SQL for $F$, $E$ and $M$-operators in path expansion in line 4. After that, we detect the number of affected tuples from SQL communication area of database (SQLCA) in line 5, and terminate iterations when *TVisited* is not updated. Otherwise, we finalize the node by its identifier $mid$. We then detect whether the target node has been finalized or not. Once the iterations are terminated, we recover the full path from the source node to the target node with iterative SQLs along the $p2s$ link.

**Listing 3: Auxiliary SQLs in Path Finding**
```
1://Detect termination
 Select * from TVisited where f=1 and nid=t;
2://Finalize the frontier node
 Update TVisited set f=1 where nid=mid
3://Locate the predecessor node
 Select p2s from TVisited where nid=xid;
```

There are at most $n$ iterations for the path finding in Algorithm 1 in the worst case, where $n$ is the number of nodes in the graph. In each iteration, we have 4 separate SQLs. Thus, we at most issue $4n$ SQLs in the shortest path discovery.

## 4. OPTIMIZATIONS FOR SHORTEST PATH DISCOVERY

In the last section, we describe the Dijkstra's shortest path algorithm using the generic *FEM* framework on RDB. In this section, we propose several techniques to further optimize the shortest path discovery in the *FEM* framework.

In the typical shortest path discovery, the main optimization is to reduce the search space [2, 24]. In the *FEM* framework, this corresponds to minimize the number of total visited nodes. To address this issue, the bi-directional search strategy is often employed [10] and we show it can be adopted in the *FEM* framework easily (in Section 4.1). Such an optimization can reduce the total computational cost contributed by relational operators, such as $F$, $E$, and $M$-operator, as fewer nodes need to be visited.

Another important and unique aspect of relational shortest path discovery is to optimize the query evaluation: in the RDB context, the set-at-a-time evaluation is more suitable than the node-at-a-time fashion for the same search space, since the former can enable database to fully adopt the intelligent scheduling to access disk and exploit the data loaded in the buffer, and thus to make a better evaluation plan [14, 18]. For example, let us consider the $E$-operator (the 3-rd SQL in Listing 2). If we have $n$ iterations in Algorithm 1, we will issue $n$ SQL statements for loading the edges of $n$ nodes separately. Such a node-at-a-time operation is very inefficient due to the redundant I/O cost for accessing edges of multiple nodes when they are stored in one data block. Thus, a natural strategy is to adopt batch processing for edge access, *e.g.,* loading the edges of multiple nodes altogether (set-at-a-time) in a single SQL statement. In other words, we would like to reduce the number of SQL statements to be issued in a shortest path discovery. However, the main issue is that the set-at-a-time fashion can easily lead to increase search space. For instance, the BFS strategy, although requires fewer SQL statements, can lead to a much larger search space compared to Dijkstra's algorithm.

We study optimization strategies to improve the performance for shortest path discovery in the *FEM* framework. Specifically, two (often-conflicting) goals need to be simultaneously achieved and balanced: i) reducing search space, which is a generic optimization criteria for shortest path discovery (and many other graph search tasks); and ii) promoting set-at-a-time query evaluation (batch processing) for the relational database operators. Two efficient techniques, bi-directional set Dijkstra's algorithm and SegTable index, have been introduced to improve the performance.

### 4.1 Bi-directional Set Dijkstra's Algorithm

The bi-directional set Dijkstra's algorithm attempts to address both batch data access and search space reduction in path finding. The path can be found by both forward expansions from the source node and backward expansions from the target node [10]. In addition, instead of selecting only one frontier node in each expansion, we select all non-finalized nodes with the same minimal distance as the frontier nodes. Such a strategy is RDB-friendly since more nodes have been processed in one operation and thus the redundant I/O cost for different nodes can be lowered by a better evaluation plan.

The bi-directional set Dijkstra's search can also be expressed inside the *FEM* framework with extensions to the existing approach in Algorithm 1. First, we revise the $F$-operator to locate a set of frontier nodes. Suppose that we can obtain $mind2s$ for the minimal $d2s$ of all non-finalized nodes (with their $f = 0$) with an auxiliary SQL operation. The predicate for the $F$-operator can then be changed to $d2s = mind2s$. Thus, the nodes with the same minimal $d2s$ will be selected by the $F$-operator. We also need another auxiliary SQL operation to finalize all frontier nodes with the predicate $d2s = mind2s$ after the $F$-operator. It is easy to know that such a batch node processing does not impact the correctness of path finding.

Second, the *TVisited* table should be extended and the selection of expansion direction is needed in the iterations. Besides $p2s$, $d2s$,

$f$ used in the forward searching, we also keep $p2t$ for the successor node to the target node, $d2t$ for the distance to the target node, and $b$ for the node finalization in the backward expansion similarly. As for the selection of expansion direction, we take the direction with fewer frontier nodes to reduce the intermediate nodes. The number of frontier nodes can be computed by an extra SQL on *TVisited*, or approximately represented by the number of affected tuples from SQLCA after the SQL statement for $M$-operator is evaluated.

Third, the bi-directional searching poses a different termination condition. Recall the bi-directional Dijkstra's algorithm [10]. Let $s$ and $t$ be source node and target node respectively, $l_f$ and $l_b$ be the minimal distance discovered in the latest forward and backward expansion respectively, $minCost$ be the minimal distance between $s$ and $t$ seen so far, $minCost$ is the shortest distance when $minCost \leq l_f + l_b$. As for our case, we can compute $minCost$ as the minimal sum of $d2s$ and $d2t$ for nodes in *TVisited* table, $l_f$ as the minimal $d2s$ for the latest forward expansion, and $l_b$ as the minimal $d2t$ for the latest backward expansion. With these values collected, we can determine whether the iterations can be terminated or not.

The bi-directional searching also brings new optimization strategy to prune search space, which can be depicted in the following. Such a pruning rule begins to work once one path between the source and the target node has been found.

THEOREM 1. *Let $l_f$ and $l_b$ be the minimal distance discovered in the latest forward and backward expansion respectively, $minCost$ be the minimal distance between $s$ and $t$ seen so far. Take the forward expansion as an example. For a frontier node $v$, we need not expand $v$ to node $x$, if $v.d2s + w(v,x) + l_b > minCost$.*

PROOF. We prove it by contradiction. Suppose that there exists a path $p' = s \rightsquigarrow t$ with $len(p') < minCost$ and $p'$ has the prefix sub-path from $s$ to $x$ via $v$. Since $v.d2s + w(v,x) + l_b > minCost$, the distance from $x$ to the target node $t$ is less than $l_b$ in $p'$ to achieve $len(p') < minCost$. In other words, $x$ has been finalized in the backward expansion with its distance to $t$ less than $l_b$. According to the same rule in the backward expansion, the path $p'$ with $len(p') < minCost$ has been already discovered. This contradicts that $minCost$ is the minimal distance discovered yet. Thus, $x$ can be ignored safely in the bi-directional searching. □.

Below, we analyze the number of iterations in bi-directional set Dijkstra's algorithm. We will see that the bi-directional set Dijkstra's algorithm needs fewer iterations than Algorithm 1. Additionally, the former incurs smaller search space than the latter. Both factors make the former outperforms the latter significantly in the RDB context.

THEOREM 2. *Given two nodes $s$ and $t$ in a graph $G$, the iterations in bi-directional set Dijkstra's algorithm in finding the shortest path from $s$ to $t$ are $\min(\delta(s,t)/w_{min}, n)$ in the worst case, where $\delta(s,t)$ is the shortest distance between $s$ and $t$, $w_{min}$ is the minimal edge weight in $G$, $n$ is the total number of nodes.*

PROOF. In the bi-directional set Dijkstra's algorithm, we finalize at least one node in each iteration. Just as Dijkstra's algorithm, the total number of iterations is no more than $n$. In addition, all non-finalized nodes with the same minimal distances can be finalized in one iteration in set Dijkstra's algorithm. Thus, the minimal distance finalized is at least $kw_{min}$ in the $k^{th}$ iteration, and then it takes $\delta(s,t)/w_{min}$ iterations to find the shortest distance $\delta(s,t)$ from $s$ to $t$ in the worst case. After that the iterations terminate since $l_f + l_b > minCost$, where $l_f$ and $l_b$ are distance finalized in the latest forward and backward expansion, and $minCost$ is $\delta(s,t)$. Thus, the number of iterations is no more than $\min(\delta(s,t)/w_{min}, n)$. □

## 4.2 Selective Path Expansion via SegTable

The bi-directional set Dijkstra's algorithm finds the shortest path in a set-at-a-time fashion. Although it requires far fewer operations on RDB than the basic method, the total number of operations is still very large, as shown in Theorem 2. Then, can we scan more nodes in the path expansion to further lower the number of operations on RDB? If so, how to select these nodes?

An extreme approach is to take BFS in finding the shortest path to reduce the number of operations. BFS expands all possible nodes including newly expanded nodes or the nodes with the reduced distance in each iteration. For a shortest path $p$, BFS can find $p$ with $e(p)$ iterations, where $e(p)$ is the total number of edges in $p$. Certainly, BFS requires no more operations than other methods. However, BFS is not always effective since it scans larger search space than Dijkstra's algorithm. We can know that the nodes which have been expanded still have high chances to be re-expanded in the following iterations in BFS.

In order to address these issues, we introduce an index named SegTable to preserve pre-computed shortest segments. We then can exploit these segments to select the partial nodes for further expansion. We wish to reduce the total number of iterations while not enlarging search space seriously.

**SegTable.** Intuitively, the edges with smaller weights have more chances to be in shortest paths. Then, we can pre-compute the shortest segments with their distances less than a given threshold, and store the results into SegTable. Roughly speaking, SegTable can be viewed as a shortest path index.

DEFINITION 4. **SegTable.** *Let $G = (V, E)$ be a graph, $l_{thd}$ be a threshold. SegTable includes TOutSegs and TInSegs tables. TOutSegs preserves the pre-computed segments in the outgoing direction. Each tuple (fid, tid, pid, cost) in TOutSegs can be composed by:*

*(1) $(u, v, pre(v), \delta(u,v))$, for any node pair $u, v \in V$ with $\delta(u,v) \leq l_{thd}$. Here $\delta(u,v)$ is the shortest distance between $u$ and $v$, and $pre(v)$ is the predecessor of $v$ in the shortest path from $u$ to $v$.*

*(2) $(u, v, u, w(u,v))$, for any $(u,v) \in E$ and $\delta(u,v) > l_{thd}$. Here $w(u,v)$ is the weight of the edge $(u,v)$.*

*TInSegs preserves the pre-computed segments in the incoming direction in a similar way. $l_{thd}$ is called the index threshold.*

SegTable is actually the representation for a graph $G'$ containing pre-computed segments and original edges in graph $G$. Figure 4(a) shows the graph $G'$ with segments on the original graph $G$ in Figure 1 with $l_{thd} = 6$. The edge $(s, e)$ with the weight 4 is a pre-computed segment. If we start path searching from $s$, $e$ can be found in one expansion instead of two expansions. The edge $(s, d)$ with the weight 2 is also a pre-computed segment. The refined edge weight can avoid unnecessary re-expanding from $d$ on the original graph. The edge $(e, h)$ is the original edge in $G$ with $\delta(e, h) > l_{thd}$. The relational tables for SegTable graph are shown in Figure 4(b).

**Selective Path Expansion.** Since SegTable contains the segments which have more chances to be in shortest paths, we attempt to select the partial nodes from SegTable in the expansions. Let $l_{thd}$ be the index threshold. Take the $k^{th}$ forward expansion as an example. A node $u$ is selected as a frontier node from visited nodes, if $u.d2s$ is no more than $kl_{thd}$ or $u.d2s$ is the minimal among all nodes to be expanded. In other words, we prefer selecting nodes

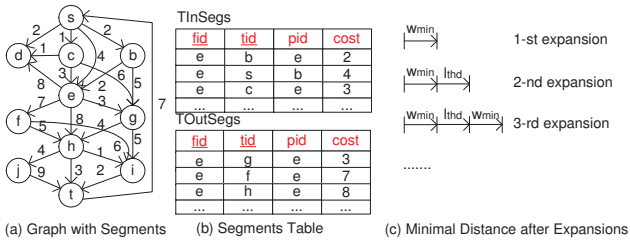

(a) Graph with Segments  (b) Segments Table  (c) Minimal Distance after Expansions

**Figure 4: SegTable with Index Threshold $l_{thd} = 6$**

with smaller $u.d2s$ since a larger $d2s$ indicates a higher chance in unnecessary re-expansions.

We extend the two-value sign $f$ for node finalization into three-value sign to support the complex rule in selecting frontier nodes. As that in Listing 2, $f = 0$ on a node $u$ indicates $u$ is a candidate frontier node, and $f = 1$ on a node $u$ means $u$ has been expanded before. Here, we introduce $f = 2$ for the selected frontier nodes. As discussed above, a node $u$ is selected as a frontier node when $u.f = 0$, and $u.d2s \leq kl_{thd}$ or $u.d2s$ is the minimal among all nodes with $f = 0$. In order to simplify the expression of $F$-operator, we can select these frontier nodes and change their sign $f$ to 2 with an auxiliary SQL before $F$-operator. Hence, $f = 2$ is the only predicate in the $F$-operator. In addition, after the path expansion, we still use $f = 2$ to distinguish the frontier nodes just now, and changes their $f$ to 1 indicating that these nodes have been expanded.

The selective path expansion over SegTable can find the shortest path with no more iterations than the set Dijkstra's algorithm on the original graph. We first present an intuitive idea in Figure 4(c) and will give a formal description in the following. Let $l_{thd}$ be the index threshold, $w_{min}$ for the minimal edge weight. The minimal distance $l_f$ found in the 1-st forward expansion is $w_{min}$ in the worst case. $l_f$ will be $l_{thd} + w_{min}$ in the 2-nd expansion, since the distance less than $l_{thd} + w_{min}$ has been discovered in the 1-st expansion. Similarly, $l_f$ in the 3-rd expansion will be $l_{thd} + 2w_{min}$. Inductively, $l_f$ in the $k_f^{th}$ expansion will be $\lfloor k_f/2 \rfloor l_{thd} + \lceil k_f/2 \rceil w_{min}$. Recall that $l_f$ in the $k_f^{th}$ expansion will be $k_f w_{min}$ in the set Dijkstra's algorithm on the original graph in the worst case. $l_f$ increases faster in terms of the number of expansions than before, which makes the termination condition $l_b + l_f > minCost$ satisfied earlier.

**Construction of SegTable.** Given an original graph $G$ and an index threshold $l_{thd}$, the construction of its SegTable needs to locate all shortest segments with their distances no more than $l_{thd}$, and to find the other remaining necessary edges.

We can exploit the *FEM* framework to locate all required shortest segments in the SegTable. Take the construction of *TOutSegs* as an example. We can put all nodes in $G$ into a visited node set initially. Just like $F$-operator in the selective path searching over SegTable, in the $k^{th}$ iteration, a node $u$ is selected into frontier nodes when $u$ is a candidate frontier node, and $u.d2s < kw_{min}$ or $u.d2s$ is the minimal among all candidate frontier nodes. Here, $w_{min}$ is for the minimal edge weight. The $E$-operator is revised to restrict the maximal distance searched no more than $l_{thd}$. We repeat expanding until the minimal distance determined is more than $l_{thd}$. It is easy to know that the number of iterations in the first step is less than $l_{thd}/w_{min}$.

In the second step, we need to combine the remaining necessary edges into the SegTable $G'$ generated after the first step. Such a task can be implemented by a *merge* statement with $G'$ as the target and the original graph $G$ as the source. For an edge $e = (u, v)$ with its weight $w(u, v)$ in $G$, $e$ can be discarded when $w(u, v)$ is no less than $\delta(u, v)$ recorded in $G'$. In other cases, $e$ with its weight $w(u, v)$ is added into $G'$.

### 4.3 Complete Algorithm of Bi-Directional Selective Path Expansion on SegTable

Next, we give the complete algorithm of bi-directional selective path searching on SegTable. We insert two given nodes into the empty *TVisited* table in line 1. $minCost$ is for the minimal distance currently discovered. The variable $l_f$ for the minimal distance in the latest expansion, $n_f$ for the number frontier nodes, and $fwd$ for the number of expansions in the forward expansion are all initialized. The counterpart variables in the backward expansion are also initialized in the same line.

---

**Algorithm 2**: Bi-directional Selective Searching on SegTable

**Input**: source node $s$ and target node $t$, SegTable.
**Output**: The shortest path between $s$ and $t$.

1. Initialize *TVisited* table with node $s$ and $t$;
2. $minCost \leftarrow +\infty$;
3. $l_f \leftarrow 0, l_b \leftarrow 0$;
4. $n_f \leftarrow 1, n_b \leftarrow 1$;
5. $fwd \leftarrow 1, bwd \leftarrow 1$;
6. **while** $l_b + l_f \leq minCost$ && $n_f > 0$ && $n_b > 0$ **do**
7.     **if** $n_f \leq n_b$ **then**
8.         Update signs for the frontier nodes with the SQL in Listing 4(1);
9.         Expand paths with the SQL in Listing 4(2);
10.         $n_f \leftarrow$ the number of affected tuples from SQLCA;
11.         Reset signs for the frontier nodes with the SQL in Listing 4(3);
12.         Locate $l_f$ with the SQL in Listing 4(4);
13.         $fwd \leftarrow fwd + 1$;
14.     **else**
15.         Similar actions from line 8 to line 13 for the backward expansion;
16.     Locate $minCost$ with the SQL in Listing 4(5);
17. Locate a node $xid$ in the shortest path with the SQL in Listing 4(6);
18. Find the sub-path $p_0$ from $s$ to $xid$ along $p2s$ links;
19. Find the sub-path $p_1$ from $xid$ to $t$ along $p2t$ links;
20. Return $p_0 + p_1$;

---

We make the path expansion in the iterations from line 6 to line 16. We select an expansion direction according to the number of frontier nodes, and show the actions in the forward expansion from line 8 to line 13. First, we set $f = 2$ on the selected frontier nodes from all candidate frontier nodes ($f = 0$). We then implement $F$, $E$, and $M$-operator with the 2-nd SQL in Listing 4. Compared to the 3-rd and 4-th SQL in Listing 2, the 2-nd SQL in Listing 4 uses pre-computed segments *TOutSegs* instead of *TEdges*, specifies frontier nodes with $f = 2$, and takes bi-directional pruning rule $out.cost + q.d2s + l_b < minCost$. Next, we update the sign $f = 1$ on the frontier nodes just now ($f = 2$) to avoid the repeat expansion in the following. After the path expansion, we obtain $l_b$ for the minimal $d2s$ in the latest forward expansion and $minCost$ for the minimal distance currently discovered to determine whether the iterations can be stopped. Once the iterations are terminated, we locate one node $xid$ in the shortest path with $minCost$, and recover the full shortest path along the $p2s$ and $p2t$ links from $xid$.

**Listing 4: SQL in Expansion**
```
1://change signs for frontiers in fwd-th expansion
  Update TVisited set f=2
    where (d2s<=fwd*lthd or
    d2s=(select min(d2s) from TVisited where f=0))
    and f=0;
2://Make fwd-th expansion with F, E, M-Operator
  Merge TVisited as target
    using (select nid, p2s, cost
      from ( select out.tid, out.pid,
              out.cost+q.d2s, row_number() over
                (partition by out.tid order by
                  out.cost+q.d2s asc) as rownum
            from TVisited q, TOutSegs out
            where q.nid=out.fid and q.f= 2 and
              out.cost+q.d2s+lb<minCost)
            tmp (nid, p2s, cost,rownum)
      where rownum=1
  ) as source(nid, p2s, cost)
  on source.nid=target.nid
  when matched and target.d2s>source.cost then
    update set d2s=source.cost,
      p2s=source.p2s, f=0
  when not matched by target then
    insert (nid, d2s, d2t, p2s, f)
      values( source.nid, cost,Max, source.p2s, 0);
3://Reset the sign that node has been expanded.
  Update TVisited set f=1 where f=2;
4://Locate the minimal distance in forward search
  Select min(d2s) from TVisited where f=0;
5://Locate the minimal distance discovered
  Select min(d2s+d2t) from TVisited;
6://Locate a node in the shortest path
  Select nid from TVisited where d2s+d2t=minCost;
```

The shortest path finding in Algorithm 2 requires no more iterations than the former methods, which can be illustrated as follows:

THEOREM 3. *Given a source node $s$ and a target node $t$, the SegTable with the index threshold $l_{thd}$, the number of iterations in Algorithm 2 is less than $min(n, \frac{2(\delta(u,v)+l_{thd})}{l_{thd}+w_{min}})$.*

PROOF. Algorithm 2 will terminate when $l_f + l_b \geq minCost$, where $l_f$ and $l_b$ are the minimal distance found in the latest forward and backward expansions respectively, and $minCost$ is the minimal distance between $s$ and $t$ seen so far. Since each iteration will finalize at least one node, the total iterations are no more than $n$ in the worst case. Let $k_f$ and $k_b$ be the number of forward and backward expansions respectively. As illustrated in Figure 4(c), $l_f$ is no less than $\lfloor k_f/2 \rfloor l_{thd} + \lceil k_f/2 \rceil w_{min}$, and $l_b$ is no less than $\lfloor k_b/2 \rfloor l_{thd} + \lceil k_b/2 \rceil w_{min}$. $l_b + l_f > ((k_b + k_f)/2)(l_{thd} + w_{min}) - l_{thd}$. Thus, when the total number of iterations, $k_b + k_f$, reaches $\frac{2(\delta(u,v)+l_{thd})}{l_{thd}+w_{min}}$, the iterations can be terminated with $l_b + l_f \geq \delta(u,v)$. □

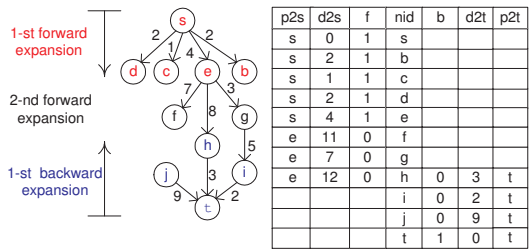

**Figure 5: TVisited in Bi-directional Searching on SegTable**

The path finding via SegTable can be illustrated in Figure 5. Let $s$ be the source node and $t$ be the target node. We make 2 forward expansions and 1 backward expansion. *TVisited* table contains the current intermediate results. The current shortest distance $minCost$ is 15, achieved on node $h$. We have to continue searching, since $l_f = 12$, $l_b = 2$, and $l_f + l_b < minCost$ at that time.

## 5. EXPERIMENTAL RESULTS

In this section, we experimentally evaluate the effectiveness and efficiency of our relational approach on both real and synthetic datasets extensively.

### 5.1 Experimental Setup

**Experiment Design.** We are interested in the following questions in experiments: i) Can the *FEM* framework support Dijkstra's implementation in RDB? What is the most expensive phase in path discovery? Do the new SQL features boost the performance remarkably? ii) Does the set-at-a-time evaluation method, such as set Dijkstra's algorithm, really impact the performance? Will the selective path expansion over SegTable index further improve the performance? If so, what is the impact of index threshold $l_{thd}$? iii) Can our method scale well on different datasets varying graph sizes, buffer sizes, different index strategies and different relational database systems? If all optimization strategies are used, what is the comparison between relational version and in-memory version?

**Implementation Details and Competitors.** We have implemented 7 approaches related to the paper. *DJ* is the relational version for the single directional Dijkstra's algorithm in Algorithm 1. *BDJ* is for the bi-directional Dijkstra's algorithm. *BSDJ* is for the bi-directional set Dijkstra's algorithm. *BBFS* is for the relational version of bi-directional breadth-first-search method. *BSEG* is for the selective path expansions on SegTable. These methods run on the client side which connects to the underlying RDB via JDBC. The competitors, *MDJ* and *MBDJ* are the in-memory versions for Dijkstra's algorithm and bi-directional Dijkstra's algorithm respectively. We implement all methods in Java with JDK 1.6 and evaluate them on 1.8GHz AMD processor running Windows server 2003. The maximal runtime memory of JVM is set to 1.5G.

In order to show the applicability of our method, we conduct experiments over two different relational database systems: one commercial database system (denoted by *DBMS-x*) and another open source database system PostgreSQL 9.0 (denoted by *PostgreSQL*).

We build indices over the relational tables for the graph, SegTable, and intermediate results. Specifically, we build clustered indices on *TEdges*($fid$), and on *TOutSegs*($fid$) (and on *TInSegs* similarly). In addition, we build a unique index on *TVisited*($nid$). $fid$ and $nid$ have the same meaning discussed above.

**Data Sets.** We use 5 graph data sets in tests, including three real graphs *DBLP*, *GoogleWeb*, *LiveJournal*, and two synthetic graphs named *Random* and *Power*. *DBLP* is extracted from a recent snapshot of DBLP dataset[3]. *GoogleWeb* and *LiveJournal* are downloaded from Stanford's data collection[4]. *Random* graphs are generated as follows. Let $n$ and $m$ be the number of nodes and edges respectively, we randomly select the source and target node for $m$ times among $n$ nodes. *Power* graph set is generated using *Barabasi Graph Generator v1.4*[5]. The weights of edges in all graphs are assigned randomly in [1,100].

Some statistics of these graphs are summarized in Table 1. As for any synthetic graph, we suffix $xNyd$ to indicate that the graph is with $x$ nodes and the average degree $y$. For example, *Random5mN3d* represents a *Random* graph with 5 million nodes and an average degree 3.

---
[3]http://dblp.uni-trier.de/xml/
[4]http://snap.stanford.edu/data/
[5]http://www.cs.ucr.edu/ ddreier/barabasi.html

| DataSet | # Nodes | # Edges |
|---|---|---|
| *DBLP* | 312,967 | 1,149,663 |
| *GoogleWeb* | 855,802 | 5,066,842 |
| *LiveJournal* | 4,847,571 | 43,110,428 |
| *RandomxmNyd* | 5m-40m | 5ym-40ym |
| *PowerxkNyd* | 20k-500k | 20yk-500yk |

**Table 1: Statistics of Graph Data Sets**

| | DJ | | BDJ | | BSDJ | |
|---|---|---|---|---|---|---|
| $|V|$ | Exps | Time | Exps | Time | Exps | Time |
| 20k | 9601 | 425 | 182 | 6.75 | 68 | 2.90 |
| 40k | | > 600 | 252 | 9.27 | 78 | 3.20 |
| 60k | | > 600 | 313 | 11.5 | 82 | 3.43 |
| 80k | | > 600 | 356 | 13.2 | 88 | 3.75 |
| 100k | | > 600 | 414 | 15.1 | 85 | 3.62 |

**Table 2: Exps(# Expansions), Time(Time:s) on Power Graphs**

## 5.2 Query Evaluation

We study the issues related to the *FEM* framework and its optimizations. The experiments will be divided into 3 sub-parts. We first verify the effectiveness of *FEM* framework and set-at-a-time optimization. We then study optimization strategies with SegTable. Finally, we make extensive studies on our method with all optimizations. In the following experiments, we randomly generate 100 shortest path queries, and report the average time cost.

**FEM Framework and Set-at-a-time Fashion.** In this sub-part, we first compare the time cost used by the single directional (DJ), the bi-directional (BDJ) and bi-directional set (BSDJ) Dijkstra's approach, then study the cost used by different phases and operators, and show the effectiveness of new features of SQL.

Figure 6(a) reports the results of DJ, BDJ, and BSDJ on *Power* graphs varying size from 20k to 100k. Take the shortest path discovery in a graph with 20k nodes as an example, DJ consumes about 7 minutes, while BDJ consumes 6.75 seconds, and BSDJ costs 2.25 seconds. In fact, we cannot use large graphs in this test, since DJ shows a poor performance in the shortest path discovery. Another observation in Figure 6(a) is that the time cost of BSDJ is about 1/3 of that in BDJ.

In order to find the reasons behind the performance improvement, we collect the total number of expansions in path finding and the time cost consumed by DJ, BDJ, and BSDJ in Table 2. It clearly shows that BSDJ takes the fewest expansions. The number of expansions in DJ is about 50 times bigger than that in BDJ, and 140 times bigger than that in BSDJ. Thus, the SQLs used by BSDJ is fewest among three. The results verify our claim before. The set-at-a-time evaluation fashion, which enables the RDB optimizer to produce a better evaluation plan, can beat the node-at-a-time evaluation fashion easily, when they have the same search space.

Figure 6(b) plots the time cost used in different phases in the path finding in Algorithm 1, including the path expansion (denoted by *PE*), the statistics collection (denoted by *SC*), and the full path recovery (denoted by *FPR*). We can see that the path expansion with three operators consumes most of time. Next, we go deep into the path expansion. We directly translate $F$, $E$, and $M$-operators into separate SQLs, and collect their time cost. The results are presented in Figure 6(c). We find $E$-operator takes about 75 percent time in the shortest path discovery. It is because $E$-operator makes a join with the graph table to find newly expanded nodes.

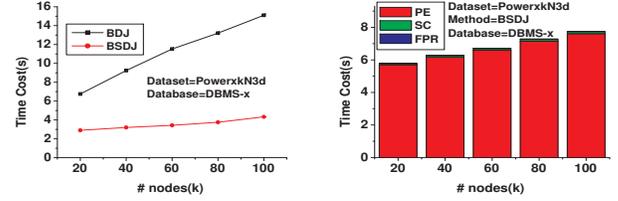

(a) Query time vs. graph scale  (b) Query time vs. phases

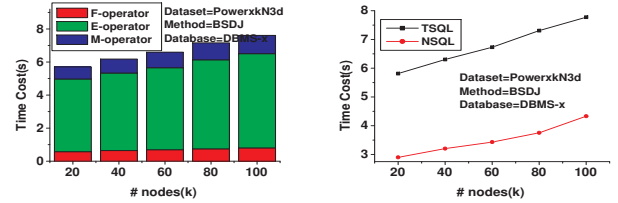

(c) Query time vs. operators  (d) Query time vs. SQL

**Figure 6: Experimental Results on FEM Framework**

Figure 6(d) studies the query time cost affected by different SQL features. We implement path discovery with the new features including window function and merge statement (denoted by *NSQL*) and the traditional features including aggregate functions and insert/update for merge (denoted by *TSQL*). The results clearly show that the *NSQL* method outperforms the *TSQL* method significantly. We believe that the advance of RDB makes it more powerful in handling the shortest path discovery and other complex operations.

Due to the fact that BSDJ outperforms DJ and BDJ thoroughly, we report only the curve for BSDJ, and omit the curves for the other two in the following. We will use new features of SQL in path finding unless explicitly mentioned.

**Optimizations using SegTable.** Next, we analyze the performance of the selective path expansion on SegTable (BSEG). The term BSEG($x$) is for BSEG with the index threshold $l_{thd} = x$. In order to see the trade-off between the search space and the evaluation fashion clearly, we also implement BBFS for bi-directional breadth-first-search.

Figure 7(a) and Figure 7(b) compare the time cost consumed by BSDJ, BBFS and BSEG on two sets of large graphs. BSEG is fastest among three. The time cost in BSEG is nearly 1/7 of that in BBFS and 1/3 in BSDJ across *LiveJournal4m* graph, and the proportion is about 1/2 or 1/3 on *Random* graphs. We make a deep analysis on the time cost, the number of expansions and the size of visited nodes in BSDJ, BBFS and BSEG. The results are listed in Table 3. We can see that although the visited nodes in BSEG are a little more than those in BSDJ, the number of expansions in BSEG is just 1/3 of that in BSDJ. The reduction in the corresponding SQLs at the cost of slightly enlarged search space makes BSEG work best. As for BBFS, it takes fewest expansions in the tests. However, BBFS is slower than other two methods in some cases since BBFS incurs more visited nodes. By combining these experiment results, we know that we cannot consider the evaluation fashion (the number of operations) or the search space (the size of intermediate nodes) separately, but strive to achieve a balance between them.

Figure 7(c) and Figure 7(d) illustrate the query evaluation time cost varying the index threshold $l_{thd}$ of SegTable on both *Power* graph and real data respectively. We observe that the performance is improved first and then declines with the increase of the index threshold $l_{thd}$. We explain the reasons as follows: on the one hand, according to our shortest path discovery algorithm, a larger $l_{thd}$ results in more segments in SegTable and fewer expansions to locate the shortest path. On the other hand, more pre-computed segments enlarge the search space, which cuts down the query performance. In addition, as shown in Figure 7(c) and Figure 7(d), a relatively large $l_{thd}$ (*e.g.*, $l_{thd}$=30) is appropriate for *Power* graph while a smaller $l_{thd}$ (*e.g.*, $l_{thd}$=6 or 8) is more suitable on our real graphs. How to find an optimal $l_{thd}$ for SegTable over different graphs will be a future work of this paper.

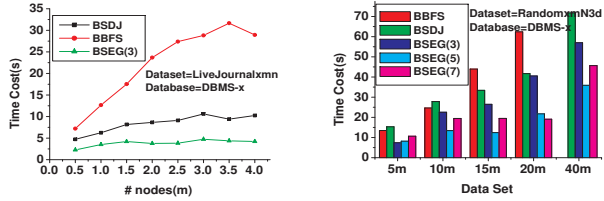

(a) Query time vs. graph scale  (b) Query time vs. graph scale

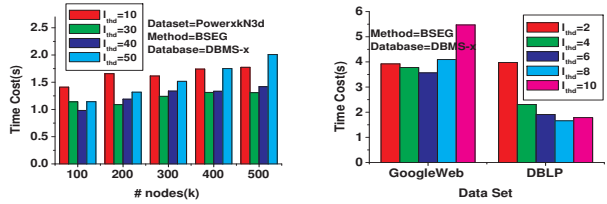

(c) Query time vs. $l_{thd}$  (d) Query time vs. $l_{thd}$

**Figure 7: Experimental Results on SegTable**

| | BSDJ | | | BBFS | | | BSEG(5) | | |
|---|---|---|---|---|---|---|---|---|---|
| $|V|$ | Time | Exps | Vst | Time | Exps | Vst | Time | Exps | Vst |
| 5M | 15.3 | 174 | 3.6k | 13.5 | 30 | 129k | 8.2 | 57 | 4.4k |
| 10M | 27.8 | 184 | 4.9k | 24.7 | 32 | 222k | 13.4 | 60 | 7.4k |
| 15M | 33.4 | 191 | 5.9k | 44.0 | 33 | 283k | 12.4 | 61 | 9.3k |
| 20M | 41.7 | 197 | 7.4k | 62.4 | 34 | 358k | 21.7 | 62 | 10.1k |

**Table 3: Time(Time:s), Exps(# expansions), and Vst (# visited nodes) on Random Graphs**

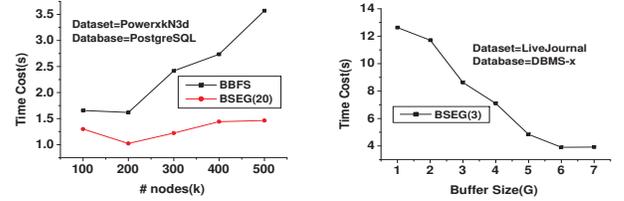

(a) Query time vs. database  (b) Query time vs. buffer size

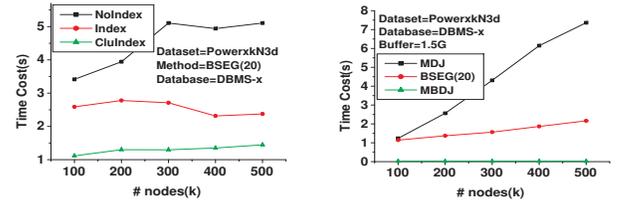

(c) Query time vs. index  (d) Query time vs. in-memory methods

**Figure 8: Experimental Results on Extensive Study**

**Extensive Studies.** In this sub-part, we study the impacts of different database engines, different buffer sizes and index strategies on the query performance. Finally, we compare our RDB approach with the in-memory ones.

Figure 8(a) compares the query time consumed by BBFS and BSEG on *PostgreSQL*. Since *PostgreSQL* supports the window function but cannot provide the *merge* statement, we use *insert* and *update* statement for the $M$-operator instead. The results on *PostgreSQL* are similar to those on the commercial database system, which also shows that our method has good applicability.

Figure 8(b) illustrates the impact of different buffer sizes in RDB on the query performance. It is not surprised that the increase of the buffer size decreases the number of I/O operations, and then improves the performance. For our case, the decrease of cost is nearly linear to the increase of the buffer sizes. We also notice that once the buffer size reaches a threshold, for example, 6G for the shortest path discovery over *LiveJournal*, the evaluation time stays almost stable. In such a setting, the graph can be fully loaded into memory, and the former key factor in I/O operations is not as important as before.

In Figure 8(c), we conduct studies on different index strategies, including the clustered and unique index (denoted by *CluIndex*), non-clustered unique index (denoted by *Index*), and no index (denoted by *NoIndex*) on *TOutSegs*($fid$) (on *TInSegs* similarly), as well as on *TVisited*($nid$). We can see that *CluIndex* achieves the best performance. We think that the unique index may improve the performance of $E$-operator, which can support the index-join between the SegTable and *TVisited* table, and the clustered index can reduce I/O cost in accessing edges for given nodes.

Last, we compare our relational approach with the in-memory ones, including Dijkstra's algorithm (MDJ) and bi-directional Dijkstra's algorithm (MBDJ). We set both the buffer size of RDB and the maximal memory used in in-memory approaches to 1.5G. In order to make the comparison fair, the time reported for in-memory approaches does not include the time in loading graph into memory, and the time for relational approaches is collected after the database buffer becomes hot. We can see our BSEG algorithm is not as fast as MBDJ in Figure 8(d). It is expected as RDB is a general infrastructure to different information systems, and is not tailored for the graph data management. In addition, the buffer of RDB also contains other information such as system tables. We also notice that BSEG can outperforms MDJ and shows better scalability. These results also reveal that the proper evaluation strategy in the RDB context still can gain a satisfactory performance. We stress here that the main advantage of RDB lies in its scalability, stability and easy programming in the graph management.

## 5.3 SegTable Construction

In this part, we study the construction of SegTable for two reasons. One the one hand, SegTable is an important optimization strategy in the path discovery. On the other hand, the construction of SegTable can be viewed as another application of the *FEM* framework. These tests can be the supplement to the query evaluation ones.

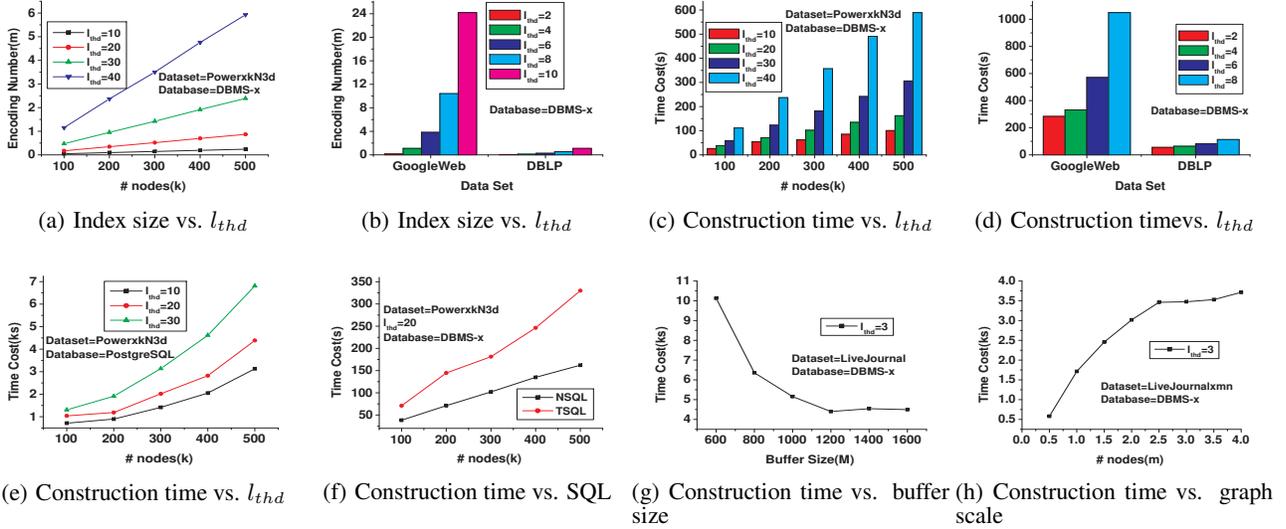

(a) Index size vs. $l_{thd}$  (b) Index size vs. $l_{thd}$  (c) Construction time vs. $l_{thd}$  (d) Construction time vs. $l_{thd}$

(e) Construction time vs. $l_{thd}$  (f) Construction time vs. SQL  (g) Construction time vs. buffer size  (h) Construction time vs. graph scale

**Figure 9: Experimental Results on SegTable Construction**

**Index Size.** Figure 9(a) and Figure 9(b) plot the index size varying the index threshold $l_{thd}$ on *Power* graphs and real datasets respectively. We can know that a larger $l_{thd}$ requires pre-computing more segments, which results in a larger SegTable index. We also observe that, for different graphs, the impact of $l_{thd}$ on index size varies. For example, *GoogleWeb* is more sensitive to $l_{thd}$ than *DBLP*. It is partially due to that *GoogleWeb* has a skewed degree distribution.

**Construction Time.** Figure 9(c) and Figure 9(d) summarize the construction time cost varying the index threshold $l_{thd}$ across both synthetic and real datasets on commercial database. The results and reason are similar to those in the tests on the index size. A larger $l_{thd}$ corresponds to longer shortest segments and then incurs more indexing time.

Figure 9(e) shows the time results on open source database *PostgreSQL* across *Power* graphs. The behavior on *PostgreSQL* is similar to that on the commercial database *DBMS-x*, such as Figure 9(c) and Figure 9(d). This proves our SegTable based method can be applied on different relational database platforms.

Figure 9(f) compares the construction time cost using the new and traditional features of SQL on *Power* graphs. Since the intermediate nodes in the index construction are also restricted by the index threshold $l_{thd}$, the advantage of the method using *NSQL* over the method using *TSQL* is not as significant as that in the path finding. However, we can see the method using *NSQL* still outperforms the one using *TSQL*.

Figure 9(g) illustrates the impact of different buffer sizes used by RDB on the index construction time cost. The results are similar to those in the query evaluation. The increase of buffer size will improve the indexing performance. For example, the indexing time cost with a memory set to 0.6G is two times of that with memory fixed to 1.6G. Moreover, when the buffer size exceeds 1.2G, the maximal requirement on memory for indexing, the time curve is almost horizonal.

Figure 9(h) shows the index construction time varying the size of graphs. A larger graph needs more visited nodes in the construction, which consumes more indexing time. Additionally, we observe that our construction method has high scalability as graphs grow larger, as the relationships between the index construction time and graph size are almost linear. It is due to that our SegTable index only encodes local shortest segments.

## 5.4 Summary

To sum up, from the experimental results, we can draw the following conclusions: i) Our *FEM* framework can be used to answer the graph search queries such as the shortest path discovery. And the new features introduced by recent SQL standards can improve the performance of *FEM* greatly. ii) The set-at-a-time evaluation method, such as BSDJ, can outperform the node-at-a-time evaluation method, such as BDJ, significantly. iii) The index, SegTable, can further improve the performance by reducing the number of expansions at the cost of slightly enlarged search space, if the index threshold is properly set. iv) Our relational approach shows high scalability in terms of graph sizes, buffer sizes, and different database management systems.

## 6. RELATED WORK

Graph search queries are very basic and important graph operations. Due to the large search space, many proposed methods take greedy approaches, including Dijkstra's algorithm for the shortest path [12], Prim's algorithm for the minimal spanning tree [20], the greedy salesman path discovery [9], and the like.

Shortest path discovery is a representative graph search operation. Dijkstra's algorithm is a well-known online algorithm [12] to solve a single-source shortest path problem in $O(n^2)$ time on general graphs. The bi-directional search strategy [10] is an important extension to reduce the search space. The shortest paths can be pre-computed and stored by different forms, such as the landmark index [19, 2], 2-HOP related index [11], TEDI index [24], etc. These indices can improve the performance in the running time. However, these online shortest path discovery and shortest index building methods do not consider the case when the graph cannot be fully loaded into the memory.

The external graph operations are needed with the rapid growth of graphs. An external shortest path index is proposed by [8] on planar graphs. MapReduce framework [16, 3] and its open source implementation Hadoop [1] achieve high scalability in handling large graphs. The current limitation of MapReduce framework lies in

368

its weak support to online query and expensive dynamic update of graphs. We also notice specific graph operations in the external memory. For example, the approximate minimum-cut [5] can be computed on the sampled graph. The cliques can be found on the partially loaded sub-graphs [15]. However, it is hard to extend them to support general graph search queries.

The extension of RDB as a scalable platform to manage complex data types or support sophisticated applications has been a hot topic recently. The data mining tasks [4] and statistic data analysis [6] can be evaluated on the top of RDB. When using RDB to manage XML data [17, 13, 18], an XML query is always translated into multiple SQLs, during which an important optimization is to produce fewer SQLs and thus the optimizer in RDB can generate better evaluation plans [18].

RDB based graph query processing is closely related to our work. The reachability query is evaluated by a stored procedure [23]. The SQL based graph data mining has been studied in [22, 7]. Different from the existing methods, we attempt to design a generic graph search framework in RDB, and improve its performance by exploiting new features of SQL. We also design optimizations to further improve the performance by balancing the search space reduction and RDB-friendly evaluation fashion.

## 7. CONCLUSIONS AND FUTURE WORK

This paper focuses on the relational approach to discover the shortest path on large graphs. We abstract a relational generic graph search framework *FEM* with three new operators, and employ the new features of SQL such as window function and merge statement to improve the performance of the framework. Additionally, we design two optimizations for shortest path discovery inside the framework, including the bi-directional set Dijkstra's searching and selective path expansion on SegTable containing pre-computed shortest segments. The final experimental results show the scalability and efficiency of our relational approach.

This work can be extended in several interesting directions. First, we will study the evaluation of other graph search queries, such as graph pattern match, using the *FEM* framework. Second, we will exploit the distributed database to achieve higher scalability in terms of graph sizes. The partition of the relational tables for graphs and intermediate results among distributed database is an interesting issue. Third, the relational approach to graph management needs to consider the dynamic changes of the original graph. The pre-computed results, such as SegTable in this paper, should be maintained incrementally.

## ACKNOWLEDGMENTS

We would like to thank the anonymous reviewers for their helpful comments. The NSFC supported Gao via 60873062 and 61073018, and supported Jin via 61003167. The research grants Council of the Hong Kong SAR supported Yu via 419008 and 419109. The NSF supported Jin via IIS-0953950. National science and technology major program supported Wang via 2010ZX01042-001-003-05 and 2010ZX01042-002-002-02.

## 8. REFERENCES

[1] *Apache Hadoop*. http://hadoop.apache.org.
[2] A.Goldberg and C.Harrelson. Computing the shortest path: search meets graph theory. In *SODA*, pages 156–165, 2005.
[3] B.Bahmani, K.Chakrabarti, and D.Xin. Fast personalized pagerank on mapreduce. In *SIGMOD*, pages 973–984, 2011.
[4] B.Zou, X.Ma, B.Kemme, G.Newton, and D.Precup. Data mining using relational database management systems. In *PAKDD*, pages 657–667, 2006.
[5] C.Aggarwal, Y.Xie, and P.Yu. Gconnect: A connectivity index for massive disk-resident graphs. *PVLDB*, 2(1):862–873, 2009.
[6] C.Mayfield, J.Neville, and S.Prabhakar. Eracer: a database approach for statistical inference and data cleaning. In *SIGMOD*, pages 75–86, 2010.
[7] C.Wang, W.Wang, J.Pei, Y.Zhu, and B.Shi. Scalable mining of large disk-based graph databases. In *SIGKDD*, pages 316–325, 2004.
[8] D.Hutchinson, A.Maheshwari, and N.Zeh. An external memory data structure for shortest path queries. *Discrete Applied Mathematics*, 126(1):55–82, 2003.
[9] D.Johnson and L.McGeoch. The traveling salesman problem: A case study in local optimization. *Local search in combinatorial optimization*, 215-310, 1997.
[10] D.Wagner and T.Willhalm. Speed-up techniques for shortest-path computations. In *STACS*, pages 23–36, 2007.
[11] E.Cohen, E.Halperin, H.Kaplan, and U.Zwick. Reachability and distance queries via 2-hop labels. In *SODA*, pages 937–946, 2002.
[12] E.Dijkstra. A note on two problems in connexion with graphs. *Numerische Mathematik*, pages 269–271, 1959.
[13] F.Tian, B.Reinwald, H.Pirahesh, T.Mayr, and J.Myllymaki. Implementing a scalable xml publish/subscribe system using a relational database system. In *SIGMOD*, pages 479–490, 2004.
[14] H.Garcia-Molina, J.Ullman, and J.Widom. *Database Systems: The Complete Book*. Prentice Hall Press, 2008.
[15] J.Cheng, Y.Ke, A.W.Fu, J.X.Yu, and L.Zhu. Finding maximal cliques in massive networks by h*-graph. In *SIGMOD*, pages 447–458, 2010.
[16] J.Dean and S.Ghemawat. Mapreduce: Simplified data processing on large clusters. In *OSDI*, pages 137–150, 2004.
[17] J.Shanmugasundaram, K.Tufte, C.Zhang, G.He, D.DeWitt, and J.Naughton. Relational databases for querying xml documents: Limitations and opportunities. In *VLDB*, pages 302–314, 1999.
[18] M.Benedikt, C.Chan, W.Fan, R.Rastogi, S.Zheng, and A.Zhou. Dtd-directed publishing with attribute translation grammars. In *VLDB*, pages 838–849, 2002.
[19] M.Potamias, F.Bonchi, C.Castillo, and A.Gionis. Fast shortest path distance estimation in large networks. In *CIKM*, pages 453–470, 2009.
[20] R.Prim. Shortest connection networks and some generalizations. *Bell System Technical Journal*, 36:1389–1401, 1957.
[21] R.Ronen and O.Shmueli. Soql: A language for querying and creating data in social networks. In *ICDE*, pages 1595–1602, 2009.
[22] S.Srihari, S.Chandrashekar, and S.Parthasarathy. A framework for sql-based mining of large graphs on relational databases. In *PAKDD*, pages 160–167, 2010.
[23] S.Trißl and U.Leser. Fast and practical indexing and querying of very large graphs. In *SIGMOD*, pages 845–856, 2007.
[24] F. Wei. Tedi: efficient shortest path query answering on graphs. In *SIGMOD*, pages 99–110, 2010.
[25] W.Fan, J.Li, S.Ma, N.Tang, Y.Wu, and Y.Wu. Graph pattern matching: From intractable to polynomial time. *PVLDB*, 3(1):264-275, 2010.